\documentclass[12pt]{article}
\usepackage{amsthm}
\usepackage{amsmath,amssymb,graphicx,latexsym}
\usepackage{epsfig}
\usepackage{calc}
\usepackage{setspace}
\usepackage[noadjust]{cite}

\newtheorem{theorem}{Theorem}[section]

\newtheorem{corollary}{Corollary}[section]

\newtheorem{definition}{Definition}[section]

\newtheorem{lemma}{Lemma}[section]

\newtheorem{remark}{Remark}[section]
\def\R{{\mathbb{R}}}
\def\X{{\mathbb{X}}}
\def\B{{\mathcal{B}}}
\def\D{{\mathcal{AC}}}
\def\P{{\mathcal{P}}}
\def\H{{\mathcal{H}}}
\def\h{{\mathbb{H}}}
\def\d{{\mathcal{D}}}
\def\M{{\mathcal{M}}}
\def\I{{\mathcal{I}}}

\DeclareMathOperator{\esssup}{ess\,sup}
\DeclareMathOperator{\supp}{support}
\newenvironment{keywords}{\begin{center} \begin{small} \textbf{Index Terms.}}{\end{small}\end{center}}
\begin{document}
%
\title{On Convergence Properties of Shannon Entropy}

\author{Francisco J. Piera\thanks{F.J. Piera is with the Department of Electrical Engineering, University of Chile, Av. Tupper 2007, Santiago, Casilla 412-3,
Chile (e-mail: fpiera@ing.uchile.cl).}
\and Patricio Parada
\thanks{P. Parada is with the Department of Electrical and Computer Engineering, University of Illinois at
Urbana-Champaign, 1406 W. Green St., Urbana, IL, 61801-2918 USA,
and the Department of Electrical Engineering, University of Chile,
Av. Tupper 2007, Santiago, Casilla 412-3, Chile (e-mail:
paradasa@uiuc.edu).}}
\date{December 27, 2006}
\maketitle

\begin{abstract}
Convergence properties of Shannon Entropy are studied. In the
differential setting, it is shown that weak convergence of
probability measures, or convergence in distribution, is not
enough for convergence of the associated differential entropies. A
general result for the desired differential entropy convergence is
provided, taking into account both compactly and uncompactly
supported densities. Convergence of differential entropy is also
characterized in terms of the Kullback-Liebler discriminant for
densities with fairly general supports, and it is shown that
convergence in variation of probability measures guarantees such
convergence under an appropriate boundedness condition on the
densities involved. Results for the discrete setting are also
provided, allowing for infinitely supported probability measures,
by taking advantage of the equivalence between weak convergence
and convergence in variation in this setting.
\end{abstract}

\begin{keywords}
Shannon entropy, continuous/discrete alphabet sources,
compactly/uncompactly supported densities, weak convergence,
convergence in variation, Kullback-Liebler discriminant
\end{keywords}


\renewcommand{\labelenumi}{(\roman{enumi})}
\newpage
\section{Introduction}\label{intro}
Convergence of a sequence of probability measure entropies plays a
key role in information theory, from both theoretical and applied
points of view, often appearing linked to the problem of estimation
of the entropy of a source \cite{WF2003,AK2002,WG1999,HM1993}.

As it is usual in information theory, the first order of business
is to understand the problem in the context of discrete sources,
and some of the convergence results can be found in today's
standard textbooks of the area \cite{RB1987,TC1991}, and some
recent works \cite{SWH2005}. A more general approach can be found
in the works of A. Barron, where a proof of the Central Limit
Theorem based on entropy convergence \cite{AB1986} and the entropy
convergence of stationary processes \cite{AB1985} are presented.
The discussion of information topologies for general sources
\cite{PH} touches tangentially the problem of convergence in a
more general setting.

However, the focus of many of these works has been on continuity
rather than convergence properties of Shannon entropy. On the one
hand, continuity properties embrace results guaranteeing
convergence of entropy for all approximating sequences of
probability measures converging, in a certain topology, to a given
limiting probability measure. Emphasis is put there in identifying
the largest class of probability measures for which the
corresponding convergence of entropy takes place for all
approximating sequences. On the other hand, convergence properties
are usually related to deciding whether convergence of entropy
takes place for a given, fixed family of probability measures,
also converging in a certain topology to a limiting probability
measure. Whereas in the continuity context all requirements are
imposed on the limiting probability measure, in order to ensure
convergence of entropy for all possible approximating sequences,
in the pure convergence context one can and should exploit any
underlying structure of the particular approximating sequence at
hand, as usually done in applied probability problems.

The purpose of this paper is to present general conditions for the
convergence of entropy sequences associated to both discrete and
continuous sources, over possibly infinite or non-compactly
supported alphabets, respectively.

In the case of continuous sources, results of this kind can be
used in applications where one is confronted with the problem of
deciding whether the sequence of differential entropies associated
with a family of probability densities $\{p_n\}_{n=1}^\infty$ on
$\R^k$, each term of the sequence given by
\begin{equation}\label{pri}
    -\int_{\R^k}p_{n}\log\left[p_{n}\right]dx
\end{equation}
with $dx$ denoting Lebesgue measure, converges as $n$ increases to
infinity to the respective differential entropy associated to the
limiting density of the family (assuming such limiting density
exists in some appropriate sense).

In general, only numerical computation of the sequence elements
(\ref{pri}) is possible, making it difficult to conclude the
desired convergence in an abstract sense. Such convergence must be
established then by exploiting underlying properties or structures
of the sequence $\{p_n\}_{n=1}^\infty$ by itself and its limit.

If we assume pointwise convergence of the corresponding
integrands, two main convergence-related results from real
analysis are at our disposal: the monotone and dominated
convergence theorems for Lebesgue integrals. On the one hand, the
monotone convergence theorem provides no help for this problem
given that if each $p_n$ is a probability density function and, as
such, satisfies the normalization condition
\begin{equation*}
\int_{\R^k}p_ndx=1,
\end{equation*}
then the monotonicity in the sequence $\{p_n\}_{n=1}^\infty$ is
only possible in the trivial case when all densities coincide for
almost every $x$. On the other hand, the dominated convergence
theorem requires the construction of a function $f$ such that
\begin{equation}\label{seg}
    \left|p_n(x)\log\left[p_n(x)\right]\right|\leq f(x),
\end{equation}
for each $n$ and $x$, and
\begin{equation}\label{ii}
    \int_{\R^k}fdx<\infty,
\end{equation}
being in general such construction difficult to carry out.

Though it is usually easier to check, rather than (\ref{seg}) and
(\ref{ii}), whether the boundedness condition
\begin{equation*}
    \sup_{n,x}\left|p_n(x)\right|<\infty
\end{equation*}
holds, implying then
$M\doteq\sup_{n,x}|p_n(x)\log[p_n(x)]|<\infty$, such a condition
is not enough for the application of the dominated convergence
theorem in the case of densities supported over an infinite
Lebesgue measure set, since $f$ cannot be taken as the constant
function $M(>0)$ in that case ($\int_{\X}Mdx=M\int_{\X}dx=\infty$
if $\X$ has infinite Lebesgue measure). We show, however, that
appropriate absolute continuity properties of measures provide a
suitable boundedness condition that can be used, in conjunction
with the dominated convergence theorem, to establish the desired
convergence of the associated differential entropies, and the
Kullback-Liebler discriminant as well, for densities with fairly
general supports. Our result holds independently of the
non-compact, or even infinite Lebesgue measure nature of the
supports involved. This is accomplished by exploiting the fact
that for a density $p$ on $\X\subseteq\R^k$, though Lebesgue
measure in $\X$ may be infinite if $\X$ is unbounded,
$\mu(\cdot)\doteq\int_{\cdot}pdx$ is not. The value of the result
lies on the fact that it does not require the construction of any
additional function (such as $f$ above), as it relies exclusively
on the structure of the densities involved. We also show that
convergence in distribution of the respective probability measures
is not enough to have convergence of the corresponding
differential entropies, which reinforces the importance of
establishing general conditions for such convergence to take
place.

The paper also provides a characterization of convergence of
differential entropies in terms of the Kullback-Liebler
discriminant, for densities with fairly general supports too.
Moreover, we show that under an appropriate boundedness condition
on the densities involved, convergence in variation of probability
measures does indeed guarantee the desired differential entropy
convergence.

In the discrete setting, the paper shows that convergence in
distribution and in variation of probability measures are
equivalent. In particular, if the probability measures have finite
support then convergence of their respective entropies and the
Kullback-Liebler discriminant follow immediately. In the case of
probability mass functions with infinite supports, we exploit the
afore mentioned equivalence between weak convergence and
convergence in variation to establish the convergence of entropies
and the Kullback-Liebler discriminant.

The organization of the paper is as follows. In Section
\ref{preliminary} we introduce notational and terminological
conventions used throughout the paper, as well as the necessary
elements from the theory of convergence of probability measures.
(Most of the definitions in this section apply to both the
continuous and discrete case, when Lebesgue measure does not play
a role.) Sections \ref{variation} and \ref{general} consider the
case of continuous random variables. In Section \ref{variation} we
show that convergence in distribution of the underlying
probability measures is not enough to have convergence of the
associated differential entropies, characterizing such convergence
for densities with fairly general supports in terms of the
Kullback-Liebler discriminant and showing that, under an
appropriate boundedness condition on the densities involved,
convergence in variation of probability measures does guarantee
the desired differential entropy convergence. In Section
\ref{general} we provide a general result for convergence of
differential entropy and Kullback-Liebler discriminant under a
pointwise convergence condition, taking into account both
compactly and uncompactly supported densities. In section
\ref{discrete} we deal with the discrete case. Finally, in Section
\ref{conclusions} we present a summary of the results and discuss
on their scope.

\section{Preliminary Elements}
\label{preliminary}

In this section we introduce the concepts (and related notation)
upon which we elaborate the present work. Our presentation
includes the notions of weak convergence, convergence in variation
and a measure-theoretic definition of entropy of probability
measures.

\subsection{Definitions}
Let $k$ be a positive integer, $\R^k$ the $k$-dimensional
Euclidian space endowed with the usual Euclidian metric
$\|\cdot-\cdot\|_2$, and $\B(\R^k)$ the collection of Borel sets
in $\R^k$. Also, let $\X\in\B(\R^k)$, $\X$ closed, be a Polish
subspace, i.e., $\X$ is separable (it has a countable
dense subset) and complete (every Cauchy sequence in $\X$ converges to a
point $x\in\X$) \cite{FG1997,OK2002}. We denote
as $\D(\X)$ the collection of all probability measures $\mu$ on
$(\X,\B(\X))$ which are absolutely continuous with respect to
(w.r.t.) the Lebesgue measure in $\X$ (denoted as $dx$), i.e., having
the representation
\begin{equation}\nonumber
    \mu(A)=\int_A\frac{d\mu}{dx}dx,
\end{equation}
$A\in\B(\X)$, with
$\frac{d\mu}{dx}:\X\rightarrow\R_+\doteq[0,\infty)$, Borel
measurable, the Radon-Nikodym derivative or density of $\mu$
w.r.t. $dx$. Of course, when considering $\D(\X)$ we assume $\X$
is such that $\D(\X)\neq\emptyset$ (i.e., $\X$ having strictly
positive Lebesgue measure). In the same way, we denote as
$\D_+(\X)$ the set of all $\mu\in\D(\X)$ for which
$\frac{d\mu}{dx}>0$ Lebesgue-almost everywhere on $\X$. In particular,
$\mu\in\D_+(\X)$ implies that $\mu$ and $dx$ are mutually
absolutely continuous or equivalent, and that
\begin{equation}\label{invder}
    \frac{dx}{d\mu}(x)\doteq \begin{cases}
                        \left[\dfrac{d\mu}{dx}(x)\right]^{-1} & x\in\X,\; \frac{d\mu}{dx}(x)>0 \\
                        \alpha & x\in\X,\; \frac{d\mu}{dx}(x)=0
\end{cases}
\end{equation}
with $\alpha\in\R_+$ any constant value, provides indeed a valid
expression for the Radon-Nikodym derivative $\frac{dx}{d\mu}$
(since $\mu\in\D_+(\X)$, the set $\{x\in\X:\frac{d\mu}{dx}(x)=0$\}
is Lebesgue-null).

In addition, let $f:\X\rightarrow\R$ be a real-valued function.
Its support is the closure of the set of all $x\in\X$ where $f(x)$
is strictly positive, i.e.,
\begin{equation}\nonumber
    \supp\left(f\right)\doteq\overline{\left\{x\in\X:f(x)>0\right\}},
\end{equation}
the overline $\overline{\{\cdot\}}$ denoting closure. In
particular, we have that the Lebesgue measure of the sets
$\supp(\frac{d\mu}{dx})$ and $\X$ coincide when $\mu\in\D_+(\X)$.

\subsection{Convergence of probability measures}

We now collect some basic definitions and results, in the context
needed for the next sections of the paper. Throughout, $\P(\X)$
denotes the collection of all probability measures on
$(\X,\B(\X))$ and $\mathcal{C}(\X)$ (resp., $\mathcal{C}_b(\X)$)
the space of all continuous (resp., bounded and continuous),
real-valued functions on $\X$.

\begin{definition}\label{weakc}
A sequence $\{\mu_n\}_{n=1}^{\infty}\subseteq\P(\X)$ is said to
converge weakly to $\mu\in\P(\X)$, denoted $\mu_n\Rightarrow\mu$
as $n\uparrow\infty$, if
\begin{equation}\nonumber
    \int_{\X}fd\mu_n\rightarrow\int_{\X}fd\mu
\end{equation}
as $n\uparrow\infty$ for each $f\in\mathcal{C}_b(\X)$.
\end{definition}

Since $\X$ is separable, weak convergence $\mu_n\Rightarrow\mu$ as
$n\uparrow\infty$ of $\{\mu_n\}_{n=1}^{\infty}\subseteq\P(\X)$ to
$\mu\in\P(\X)$ is equivalent to convergence
$\rho(\mu_{n},\mu)\rightarrow 0$, as $n\uparrow\infty$ as well,
with $\rho(\cdot,\cdot)$ denoting the Prohorov metric on
$\P(\X)\times\P(\X)$, i.e.,
\begin{multline}\nonumber
    \rho(\sigma_{1},\sigma_{2})\doteq\inf\{\epsilon>0:\sigma_{1}(A)
    \leq\sigma_{2}(A^{\epsilon})+\epsilon,\\
    \sigma_{2}(A)\leq\sigma_{1}(A^{\epsilon})+\epsilon, \forall
    A\in\B(\X)\},
\end{multline}
$\sigma_{1},\sigma_{2}\in\P(\X)$, where for $A\subseteq\X$,
$\epsilon>0$ and $y\in\X$,
$A^{\epsilon}\doteq\{x\in\X:d(x,A)<\epsilon\}$ and
$d(y,A)\doteq\inf\{\|y-z\|_2:z\in A\}$. Note $A^{\epsilon}$ is
open in $\X$, and hence $A^{\epsilon}\in\B(\X)$. In addition,
since $\X$ is not just separable but Polish, $(\P(\X),\rho)$ is
Polish too \cite{PB1999}.

Weak convergence in $\P(\R^k)$ is also equivalent to the standard
convergence in distribution. (Note $\sigma\in\P(\X)$ can always be
looked at as an element of $\P(\R^k)$ by setting $\sigma(A)$ to
$\sigma(A\cap\X)$ for $A\in\B(\R^k)$.) Indeed, for
$\{\mu_{n}\}_{n=1}^{\infty}\subseteq\P(\R^{k})$ and
$\mu\in\P(\R^{k})$, we have $\mu_{n}\Rightarrow\mu$ as
$n\uparrow\infty$ if and only if, as $n\uparrow\infty$ as well,
\begin{equation}\nonumber
    F_{n}(x)\rightarrow F(x)
\end{equation}
at each $x\in\R^{k}$ point of continuity of $F$, where $F_{n}$ and
$F$ denote the distribution functions associated to $\mu_{n}$ and
$\mu$, respectively, i.e.,
\begin{align*}
    F_{n}(x)&\doteq\mu_{n}\left(\times_{i=1}^{k}\left(-\infty,x_{k}\right]\right) \\
\intertext{and}
    F(x)&\doteq\mu\left(\times_{i=1}^{k}\left(-\infty,x_{k}\right]\right)
\end{align*}
for each $x=(x_{1},\ldots,x_{k})\in\R^{k}$. In fact, the following result
holds.

\begin{lemma}\label{portmanteau}
Let $\{\mu_n\}_{n=1}^{\infty}\subseteq\P(\X)$ and $\mu\in\P(\X)$.
We have $\mu_n\Rightarrow\mu$ as $n\uparrow\infty$ if and only if,
as $n\uparrow\infty$ as well,
\begin{equation}\nonumber
    \mu_n(A)\rightarrow\mu(A)
\end{equation}
for each $A\in\B(\X)$ being a $\mu$-continuity set, i.e., such
that $\mu(\partial A)=0$ with $\partial A$ denoting the boundary
of $A$: $\partial A\doteq\{x\in\X: x\in\overline{A},x\notin A\}$.
\end{lemma}

\begin{proof}
See Portmanteau's Theorem, \cite[Theorem 2.1, p.16]{PB1999}.
\end{proof}

Another important way of convergence for probability measures,
stronger than weak convergence, is the so-called convergence in
variation associated with the distance in variation between
probability measures.

\begin{definition}
The distance in variation between $\sigma_1\in\P(\X)$ and
$\sigma_2\in\P(\X)$ is the real number
$\|\sigma_1-\sigma_2\|_{V}\in[0,2]$ given by
\begin{equation}\nonumber
    \|\sigma_1-\sigma_2\|_{V}\doteq\sup_{\substack{\psi\in\M(\X)\\|\psi|\leq\mathbf{1}}}
    \left|\int_{\X}\psi d\sigma_1-\int_{\X}\psi d\sigma_2\right|,
\end{equation}
where $\M(\X)$ denotes the collection of all $\R^*$-valued, Borel
measurable functions on $\X$,
$\R^*\doteq\R\cup\{\pm\infty\}=[-\infty,\infty]$ is the extended real line, and
$\mathbf{1}(x)\doteq 1$, $x\in\X$. In particular, we have that
$\|\cdot-\cdot\|_{V}:\P(\X)\times\P(\X)\rightarrow[0,2]$ is indeed
a metric on $\P(\X)$. Moreover, a sequence
$\{\mu_n\}_{n=1}^{\infty}\subseteq\P(\X)$ is said to converge in
variation to $\mu\in\P(\X)$ if
\begin{equation}\nonumber
    \|\mu_n-\mu\|_V\rightarrow 0
\end{equation}
as $n\uparrow\infty$.
\end{definition}

Distance in variation can alternatively be characterized as
\begin{equation}\nonumber
    \|\sigma_1-\sigma_2\|_V=2\sup_{A\in\B(\X)}|\sigma_1(A)-\sigma_2(A)|,
\end{equation}
$\sigma_1,\sigma_2\in\P(\X)$ \cite{S1986}.

As mentioned before, convergence in variation is stronger than
weak convergence. Indeed, we have the following result.

\begin{lemma}\label{VimpliesW}
Let $\{\mu_n\}_{n=1}^{\infty}\subseteq\P(\X)$ and $\mu\in\P(\X)$.
If $\|\mu_n-\mu\|_{V}\rightarrow 0$ as $n\uparrow\infty$, then
$\mu_n\Rightarrow\mu$ as $n\uparrow\infty$.
\end{lemma}

\begin{proof}
Since $\sup_{A\in\B(\X)}|\mu_n(A)-\mu(A)|\rightarrow 0$ as
$n\uparrow\infty$, we conclude $\mu_n(A)\rightarrow\mu(A)$ for all
$A\in\B(\X)$, as $n\uparrow\infty$ as well, and therefore in
particular for each $A$ being a $\mu$-continuity set. The lemma
then follows from Lemma \ref{portmanteau}.
\end{proof}

For $\mu\in\P(\X)$ and $p\in[1,\infty)$ we define
\begin{equation}\nonumber
     L^{p}(d\mu)\doteq\left\{f\in\M(\X):\left[\int_{\X}|f|^{p}d\mu\right]^{\frac{1}{p}}<\infty\right\},
\end{equation}
with the standard convention $0[\pm\infty]=0$, and the
$L^p(d\mu)$-norm of $f\in L^{p}(d\mu)$ as
\begin{equation*}
\|f\|_{L^p(d\mu)}\doteq\left[\int_{\X}|f|^{p}d\mu\right]^{\frac{1}{p}}.
\end{equation*}

For $\mu\in\P(\X)$ we denote as $L^{\infty}(d\mu)$ the
space of all functions $f\in\M(\X)$ which are bounded except
possibly on a $\mu$-null set, and define the
$L^{\infty}(d\mu)$-norm of $f\in L^{\infty}(d\mu)$ as usual, i.e.,
\begin{equation}\nonumber
    \|f\|_{L^{\infty}(d\mu)}\doteq(\mu)\esssup_{x\in\X}|f(x)|,
\end{equation}
where for $g\in\M(\X)$, $(\mu)\esssup_{x\in\X}g(x)$, the essential
supremum of $g$ w.r.t. $\mu$, is the infimum of
$\sup_{x\in\X}h(x)$ as $h$ ranges over all functions mapping $\X$
into $\R^*$ which are equal to $g$ $\mu$-almost everywhere. Thus,
for $f\in L^{\infty}(d\mu)$ we have
\begin{equation}\nonumber
    \|f\|_{L^{\infty}(d\mu)}\\=\inf\left\{M\in\R_+:\mu\left\{x\in\X:|f(x)|>M\right\}=0\right\}.
\end{equation}
(Also, the same as for $1\leq p_1<p_2<\infty$, $f\in
L^{\infty}(d\mu)$ implies $f\in L^p(d\mu)$ for each
$p\in[1,\infty)$ and $\mu\in\P(\X)$. In fact,
$(\|f\|_{L^p(d\mu)})^p\leq\|f\|_{L^{\infty}(d\mu)}$.)

\begin{remark}
For any given $\mu\in\P(\X)$, the spaces
$(L^p(d\mu),\|\cdot\|_{L^p(d\mu)})$, $p\in[1,\infty]$, become
normed linear spaces with the usual addition and scalar
multiplication of functions, and in fact Banach spaces, provided
we treat measurable functions coinciding $\mu$-almost everywhere
as equivalent \cite{R1988}.
\end{remark}

This notion is useful to determine another characterization of
distance in variation. Let $\sigma_1$ and $\sigma_2$ be two
measures in $\D(\X)$. Then \cite{S1986}
\begin{align*}
\|\sigma_1-\sigma_2\|_V&=\left\|\frac{d\sigma_1}{dx}-\frac{d\sigma_2}{dx}\right\|_{L^1(dx)} \\
                       &=\int_{\X}\left|\frac{d\sigma_1}{dx}-\frac{d\sigma_2}{dx}\right|dx.
\end{align*}
Also note that
\begin{equation*}
    \left[\frac{d\sigma_1}{dx}-\frac{d\sigma_2}{dx}\right]\in L^1(dx)
\end{equation*}
for all $\sigma_1, \sigma_2$ in $\D(\X)$. Though not explicitly
used in the paper, for $p\in(1,\infty)$,
$\{\mu_n\}_{n=1}^{\infty}\subseteq\D(\X)$ and $\mu\in\D(\X)$ such
that
$\{(\frac{d\mu_n}{dx}-\frac{d\mu}{dx})\}_{n=1}^{\infty}\subseteq
L^p(dx)$, since
\begin{equation*}
    \left\|\frac{d\mu_n}{dx}-\frac{d\mu}{dx}\right\|_{L^1(dx)}\leq\left\|\frac{d\mu_n}{dx}-\frac{d\mu}{dx}\right\|_{L^p(dx)}
\end{equation*}
for each $n\in\{1,2,\ldots\}$, we have that convergence in
$L^p(dx)$ of the corresponding densities,
$\|\frac{d\mu_n}{dx}-\frac{d\mu}{dx}\|_{L^p(dx)}\rightarrow 0$ as
$n\uparrow\infty$, implies, the same as convergence in $L^1(dx)$,
convergence $\|\mu_n-\mu\|_V\rightarrow 0$, as $n\uparrow\infty$
as well.

\subsection{Entropy}

We conclude this section by writing a general definition of
entropy of probability measures, on measure-theoretical grounds.
In the sequel all logarithms are understood to be to the base 2.

The space of measures
\begin{equation}\nonumber
    \h(\X)\doteq\left\{\mu\in\D(X):\log\left[\frac{d\mu}{dx}\right]\in
    L^1(d\mu)\right\},
\end{equation}
with the convention $\log[0]=-\infty$, represents the set of well-defined entropy measures.

\begin{definition}
The Shannon Differential Entropy, associated to the underlying
space $\X$, is the mapping $\H:\h(\X)\rightarrow\R$, assigning to
each $\mu\in\h(\X)$ the value $\H[\mu]\in\R$ given by
\begin{equation}\nonumber
   \H[\mu]\doteq-\int_{\X}\log\left[\frac{d\mu}{dx}\right]d\mu.
\end{equation}
$\H[\mu]$ is called the Shannon Differential Entropy of $\mu$.
\end{definition}

\section{Convergence of Differential Entropy: Characterization in Terms of Weak Convergence, Convergence in Variation and the Kullback-Liebler
Discriminant} \label{variation}

In this section we illustrate by means of a counterexample how
weak convergence of probability measures is not enough for
convergence of the associated differential entropies. We
characterize the desired differential entropy convergence for
fairly general supported densities in terms of the
Kullback-Leibler discriminant, also showing that under an
appropriate boundedness condition on the densities involved,
convergence in variation of the underlying probability measures
does indeed guarantee differential entropy convergence.

Consider the space $\X=[0,1]$, and define the probability measures
(taken from \cite{PB1999}) $\mu$ and $\mu_n$ in $\D([0,1])$ by
setting, for each $x\in[0,1]$ and $n\in\{1,2,\ldots\}$,
\begin{equation}\nonumber
    \frac{d\mu}{dx}(x)\doteq 1
\end{equation}
and
\begin{equation}\nonumber
    \frac{d\mu_n}{dx}(x)\doteq
    n^2\textbf{1}\left\{x\in\bigcup_{k=0}^{n-1}\left(\frac{k}{n},\frac{k}{n}+\frac{1}{n^3}\right)\right\},
\end{equation}
where, as customary for $A\subseteq\X$, $\textbf{1}\{x\in
A\}\doteq 1$ if $x\in A$ and $\textbf{1}\{x\in A\}\doteq 0$ if
$x\in\X\setminus A$, with $\X\setminus A\doteq\{x\in\X:x\notin
A\}$, the usual set-theoretic difference. Of course,
\begin{equation*}
    \mu(A)\doteq\int_Ad\mu=\int_A\frac{d\mu}{dx}dx
\end{equation*}
and
\begin{equation*}
    \mu_n(A)\doteq\int_Ad\mu_n=\int_A\frac{d\mu_n}{dx}dx
\end{equation*}
for each $A\in\B([0,1])$. Note $\mu$ is nothing but Lebesgue measure in
$[0,1]$. Also, it is easy to see that, for each
$n\in\{1,2,\ldots\}$,
\begin{equation}\nonumber
    \left|\mu_n([0,x])-\mu([0,x])\right|\leq\frac{1}{n}\left[1-\frac{1}{n^2}\right]
\end{equation}
for all $x\in[0,1]$, and therefore $\mu_n\Rightarrow\mu$ as
$n\uparrow\infty$. On the other hand, we obviously have $\mu\in
\h([0,1])$ and $\{\mu_n\}_{n=1}^{\infty}\subseteq\h([0,1])$. In
fact,
\begin{equation}\nonumber
    \H[\mu]=-\int\limits_{[0,1]}\log\left[\frac{d\mu}{dx}\right]d\mu=-\log[1]\int\limits_{[0,1]}d\mu=0
\end{equation}
and, for each $n\in\{1,2,\ldots\}$,
\begin{align*}
    \H[\mu_n]&=-\int\limits_{[0,1]}\log\left[\frac{d\mu_n}{dx}\right]d\mu_n
            -\int\limits_{[0,1]}\log\left[\frac{d\mu_n}{dx}\right]\frac{d\mu_n}{dx}dx\\  &=-n^2\log[n^2]\int\limits_{\bigcup_{k=0}^{n-1}(\frac{k}{n},\frac{k}{n}+\frac{1}{n^3})}\negthickspace \negthickspace dx\\
    &=-2\log[n],
\end{align*}
where for the last equality above we have used the fact that
Lebesgue measure of the set
$\bigcup_{k=0}^{n-1}(\frac{k}{n},\frac{k}{n}+\frac{1}{n^3})$ is
$\frac{1}{n^2}$. Hence, we have $\mu_n\Rightarrow\mu$ as
$n\uparrow\infty$, but $\H[\mu_n]\downarrow -\infty$ as
$n\uparrow\infty$, i.e., $\H[\mu_n]\downarrow -\infty\neq
0=\H[\mu]$.

The previous counterexample shows that weak convergence of
probability measures is not enough for convergence of the
respective differential entropies. It is interesting to note that
in the example, though $\mu_n\Rightarrow\mu$ as $n\uparrow\infty$,
pointwise convergence of the family of densities
$\{\frac{d\mu_n}{dx}\}_{n=1}^{\infty}$ to $\frac{d\mu}{dx}$ fails
to hold Lebesgue-almost everywhere. Indeed, as mentioned before,
we have with
$A_n\doteq\bigcup_{k=0}^{n-1}(\frac{k}{n},\frac{k}{n}+\frac{1}{n^3})$
that
\begin{equation}\nonumber
    \mu(A_n)=\frac{1}{n^2}
\end{equation}
for each $n\in\{1,2,\ldots\}$, and therefore
\begin{equation}\nonumber
    \sum_{n=1}^{\infty}\mu(A_n)=\sum_{n=1}^{\infty}\frac{1}{n^2}<\infty.
\end{equation}
Hence, by Borel Lemma, \cite[Lemma 3, p.78]{FG1997},
\begin{equation}\nonumber
    \mu\biggl(\limsup_{n\uparrow\infty}A_n\biggr)=0,
\end{equation}
where as usual,
$\limsup_{n\uparrow\infty}A_n\doteq\bigcap_{n=1}^{\infty}\bigcup_{m=n}^{\infty}A_m$.
But then, for $\mu$-almost every $x\in[0,1]$ there exists
$n_{x}\in\{1,\ldots,n\}$ such that $x\notin A_{n}$ for all
$n\in\{n_{x},n_{x}+1,\ldots\}$. Hence, we conclude
\begin{equation}\nonumber
    \frac{d\mu_n}{dx}(x)\rightarrow 0
\end{equation}
as $n\uparrow\infty$, for $\mu$-almost every $x\in[0,1]$ as well.
Thus, we have
\begin{equation*}
    \frac{d\mu_n}{dx}(x)\rightarrow 0\neq 1=\frac{d\mu}{dx}(x)
\end{equation*}
for $\mu$-almost every $x\in[0,1]$, i.e.,
pointwise convergence of $\{\frac{d\mu_n}{dx}\}_{n=1}^{\infty}$ to
$\frac{d\mu}{dx}$ fails to hold Lebesgue-almost everywhere in
$[0,1]$.

Instead of asking for an appropriate pointwise convergence
condition, as we do in the next section, we now characterize the
desired convergence $\H[\mu_n]\rightarrow\H[\mu]$ as
$n\uparrow\infty$ in terms of the Kullback-Liebler discriminant.
Some definitions are in order before establishing the result.

For $\mu\in\P(\X)$ we denote as $\D(\X||\mu)$ the set of all
$\sigma\in\P(\X)$ that are absolutely continuous w.r.t. $\mu$,
i.e., having the representation
\begin{equation}\nonumber
    \sigma(A)=\int_A\frac{d\sigma}{d\mu}d\mu,
\end{equation}
$A\in\B(\X)$, with $\frac{d\sigma}{d\mu}:\X\rightarrow\R_+$, Borel
measurable, the Radon-Nikodym derivative or density of $\sigma$
w.r.t. $\mu$. Also, we set
\begin{equation}\nonumber
    \h(\X||\mu)\doteq\left\{\sigma\in\D(\X||\mu):\log\left[\frac{d\sigma}{d\mu}\right]\in
    L^1(d\sigma)\right\}.
\end{equation}

Considering $\sigma\in\h(\X||\mu)$ and
$\X^{\sigma}_{\mu}\doteq\supp(\frac{d\sigma}{d\mu})$ we have

\begin{align*}
\int_{\X}\log\left[\frac{d\sigma}{d\mu}\right]d\sigma
    &=\int_{\X}\log\left[\frac{d\sigma}{d\mu}\right]\frac{d\sigma}{d\mu}d\mu\\
&=\int_{\X^{\sigma}_{\mu}}\log\left[\frac{d\sigma}{d\mu}\right]\frac{d\sigma}{d\mu}d\mu \\
    &=\int_{\X^{\sigma}_{\mu}}\log\left[\frac{d\sigma}{d\mu}\right]d\sigma
\end{align*}
and, by a standard application of Jensen's Inequality \cite{TC1991},
\begin{align*}
    -\int_{\X^{\sigma}_{\mu}}\log\left[\frac{d\sigma}{d\mu}\right]d\sigma
&=\int_{\X^{\sigma}_{\mu}}\log\left[\left(\frac{d\sigma}{d\mu}\right)^{-1}\right]d\sigma\\
&\leq\log\left[\int_{\X^{\sigma}_{\mu}}\left(\frac{d\sigma}{d\mu}\right)^{-1}d\sigma\right]\\
&=\log\left[\int_{\X^{\sigma}_{\mu}}\left(\frac{d\sigma}{d\mu}\right)^{-1}\frac{d\sigma}{d\mu}d\mu\right]\\
&=\log\left[\mu(\X^{\sigma}_{\mu})\right]\\
&\leq0,
\end{align*}
with equality if and only if $\frac{d\sigma}{d\mu}=\mathbf{1}$,
$\sigma$-almost everywhere (recall $\mathbf{1}(x)\doteq 1$,
$x\in\X$). Having noticed this, we make the following
definition.

\begin{definition}
The Shannon Relative Entropy, relative to $\mu\in\P(\X)$, is the
mapping $\d[\cdot||\mu]:\h(\X||\mu)\rightarrow\R_+$, assigning to
each $\sigma\in\h(\X||\mu)$ the value $\d[\sigma||\mu]\in\R_+$
given by
\begin{equation}\nonumber
   \d[\sigma||\mu]\doteq\int_{\X}\log\left[\frac{d\sigma}{d\mu}\right]d\sigma.
\end{equation}
$\d[\sigma||\mu]$ is called the Shannon Relative Entropy between
$\sigma$ and $\mu$, or the Kullback-Liebler discriminant between
$\sigma$ and $\mu$ too.
\end{definition}

The Kullback-Liebler discriminant does not constitute a distance
between probability measures: it is not symmetric and does not
satisfies the triangle inequality; indeed, $\sigma\in\h(\X||\mu)$
does not even imply $\mu\in\D(\X||\sigma)$. It is widely used as
a notion of closeness between probability measures though, mainly
because, as shown above, $\d[\sigma||\mu]\geq 0$ with equality if
and only if $\frac{d\sigma}{d\mu}=\mathbf{1}$, $\sigma$-almost
everywhere.

Before stating the result in the next theorem, we make the
following remarks.

\begin{remark}
If $\sigma\in\D(\X)$ and $\mu\in\D_+(\X)$, then
$\sigma\in\D(\X||\mu)$. In fact, we can set
\begin{equation*}
    \frac{d\sigma}{d\mu}\doteq\frac{d\sigma}{dx}\frac{dx}{d\mu}
\end{equation*}
with $\frac{dx}{d\mu}$ given by (\ref{invder}), as we do
throughout. Then, when $\sigma,\mu\in\D_+(\X)$ we have
$\sigma\in\D(\X||\mu)$ and $\mu\in\D(\X||\sigma)$, i.e., $\sigma$
and $\mu$ are mutually absolutely continuous or equivalent.
Moreover, on the (partial) converse direction, $\sigma\in\D(\X)$
if $\sigma\in\D(\X||\mu)$ and $\mu\in\D(\X)$, and we have
$\frac{d\sigma}{dx}=\frac{d\sigma}{d\mu}\frac{d\mu}{dx}$,
Lebesgue-almost everywhere, and
$\frac{d\sigma}{d\mu}=\frac{d\sigma}{dx}[\frac{d\mu}{dx}]^{-1}$,
$\mu$-almost everywhere. These facts will be used in the sequel
without any further comment.
\end{remark}

\begin{remark}\label{Pinsker}
From Pinsker's Inequality (see for example \cite{PH}), for
any $\mu\in\P(\X)$ and
$\{\mu_n\}_{n=1}^{\infty}\subseteq\h(\X||\mu)$ we have
\begin{equation}\nonumber
    \|\mu_n-\mu\|_V\leq\sqrt{2\d[\mu_n||\mu]},
\end{equation}
for each $n\in\{1,2,\ldots\}$ and therefore, convergence
$\d[\mu_n||\mu]\rightarrow 0$ as $n\uparrow\infty$ implies
convergence $\|\mu_n-\mu\|_V\rightarrow 0$, as $n\uparrow\infty$
as well.
\end{remark}

\begin{theorem}\label{thm1}
Let $\{\mu_n\}_{n=1}^{\infty}\subseteq\h(\X)$ and $\mu\in \D(\X)$
be such that $\frac{d\mu}{dx}(x)>0$, for each $x\in\X$, and
$\log[\frac{d\mu}{dx}]\in\mathcal{C}_b(\X)$. Then,
$\{\mu_n\}_{n=1}^{\infty}\subseteq\h(\X||\mu)$, $\mu\in\h(\X)$ and
the following assertions are equivalent.
\begin{enumerate}
\item $\mu_n\Rightarrow\mu$ and $\H[\mu_n]\rightarrow\H[\mu]$ as
$n\uparrow\infty$.
\item $\d[\mu_n||\mu]\rightarrow 0$ as $n\uparrow\infty$.
\end{enumerate}
\end{theorem}

\begin{proof}
Since $\log[\frac{d\mu}{dx}]$ is in particular bounded on $\X$, we
obviously have $\mu\in\h(\X)$. In addition,
\begin{equation}\label{af}
    \int_{\X}\left|\log\left[\frac{d\mu_n}{d\mu}\right]\right|d\mu_n\\
    \leq\int_{\X}\left|\log\left[\frac{d\mu_n}{dx}\right]\right|d\mu_n+
    \int_{\X}\left|\log\left[\frac{d\mu}{dx}\right]\right|d\mu_n<\infty,
\end{equation}
since also $\{\mu_n\}_{n=1}^{\infty}\subseteq\h(\X)$. In
particular, $\{\mu_n\}_{n=1}^{\infty}\subseteq\h(\X||\mu)$. Now,
from equation (\ref{af}) we may write
\begin{align}
\label{pre}
\notag
     \d[\mu_{n}||\mu]&=\int_{\X}\log\left[\frac{d\mu_{n}}{d\mu}\right]d\mu_{n}\\ &=\int_{\X}\log\left[\frac{d\mu_{n}}{dx}\right]d\mu_{n}-\int_{\X}\log\left[\frac{d\mu}{dx}\right]d\mu_{n}
\end{align}
and, since also $\mu\in\h(\X)$, from equation (\ref{pre}) we
conclude
\begin{equation}\label{main1}
    \d[\mu_{n}||\mu]=\H[\mu]-\H[\mu_n]\\
+\int_{\X}\log\left[\frac{d\mu}{dx}\right]d\mu-\int_{\X}\log\left[\frac{d\mu}{dx}\right]d\mu_n.
\end{equation}
But, since $\log[\frac{d\mu}{dx}]\in\mathcal{C}_{b}(\X)$, if
$\mu_n\Rightarrow\mu$ as $n\uparrow\infty$ we conclude that
\begin{equation}\nonumber
       \int_{\X}\log\left[\frac{d\mu}{dx}\right]d\mu_{n}\rightarrow\int_{\X}\log\left[\frac{d\mu}{dx}\right]d\mu,
\end{equation}
as $n\uparrow\infty$ as well, equation (\ref{main1}) proving then
the implication (i) $\Rightarrow$ (ii). The converse implication
(ii) $\Rightarrow$ (i) also follows from equation (\ref{main1}),
in view of Remark \ref{Pinsker} and Lemma \ref{VimpliesW}. The
theorem is then proved.
\end{proof}

\begin{remark}\label{nnfs}
Consider $\mu\in\D(\X)$ with $\frac{d\mu}{dx}(x)>0$ for each
$x\in\X$. Then, the set $\X$ does not necessarily need to be
bounded for $\log[\frac{d\mu}{dx}]$ to be bounded on $\X$. Indeed,
consider for instance the uniform distribution on any unbounded
set $\X$ ($\subseteq\R^k$, $k>1$) having finite and strictly
positive Lebesgue measure, as for example in $\R^2$
\begin{equation}\nonumber
    \X=\left\{x=(x_1,x_2)\in\R_+^2:e^{-\lambda
    x_1}\geq x_2\right\}
\end{equation}
with $\lambda\in(0,\infty)$. $\X$ so defined is an unbounded
subset of $\R^2$. However, since the Lebesgue measure of $\X$ is
$\int_{\X}dx=\lambda^{-1}\in(0,\infty)$, the uniform distribution
$\mu_0$ on $\X$ satisfies, for all $x\in\X$,
\begin{equation}\nonumber
    \log\left[\frac{d\mu_0}{dx}(x)\right]=\log\left[\left(\int_{\X}dx\right)^{-1}\right]=\log[\lambda],
\end{equation}
trivially bounded on $\X$. In the same way, the set $\X$ does not
necessarily need to be bounded for $\log[\frac{d\mu}{dx}]$ to be
an element of $L^{\infty}(dx)$.
\end{remark}

\begin{remark}
Since $\X$ is closed, we have
$\log[\frac{d\mu}{dx}]\in\mathcal{C}_b(\X)$ whenever $\X$ is in
addition bounded and $\mu\in\D(\X)$ with $\frac{d\mu}{dx}(x)>0$,
for each $x\in\X$, and $\frac{d\mu}{dx}\in\mathcal{C}(\X)$.
Indeed, if $\X\subseteq\R^{k}$ is closed and bounded it is then
compact, and therefore for $\mu\in\D(\X)$ with
$\frac{d\mu}{dx}(x)>0$, for each $x\in\X$, and
$\frac{d\mu}{dx}\in\mathcal{C}(\X)$, there exist $m$ and $M$ in
$(0,\infty)$, $m\leq M$, such that $\frac{d\mu}{dx}(x)\in[m,M]$,
for each $x\in\X$ as well. Thus,
$\log[\frac{d\mu}{dx}]\in\mathcal{C}_b(\X)$. Also, note that for
the purpose of Theorem \ref{thm1} we can always take $\X$ as being
bounded for $\{\mu_n\}_{n=1}^{\infty}\subseteq\D(\R^k)$ and
$\mu\in \D(\R^k)$ when
\begin{equation}\nonumber
    \bigcup_{n=1}^{\infty}K_n\subseteq K
\end{equation}
and $K$ is bounded, where $K\doteq\supp(\frac{d\mu}{dx})$ and
$K_n\doteq\supp(\frac{d\mu_n}{dx})$ for each $n\in\{1,2,\ldots\}$
(all the supports being taken w.r.t. $\R^k$). Indeed, with
$\X\doteq K$ we then have $\frac{d\mu}{dx}>0$ on $\X$ and each
$\mu_n$, the same as $\mu$, is concentrated on $\X$, i.e.,
$\mu_n(\X)=1$.
\end{remark}

\begin{remark}
The probability measures considered in the counterexample at the
beginning of this section satisfies all hypotheses of Theorem
\ref{thm1} and, in addition, $\mu_n\Rightarrow\mu$ as
$n\uparrow\infty$. However, similar to the differential entropy
convergence failure, we have $\d[\mu_n||\mu]=2\log[n]\uparrow
\infty$ as $n\uparrow\infty$, i.e., the convergence
$\d[\mu_n||\mu]\rightarrow 0$ as $n\uparrow\infty$ fails to hold.
Also,
\begin{equation}\nonumber
    \sup_{A\in\B(\X)}|\mu_n(A)-\mu(A)|\geq|\mu_n(A_n)-\mu(A_n)|=1-\frac{1}{n^2},
\end{equation}
for each $n\in\{1,2,\ldots\}$, and hence the convergence
$\|\mu_n-\mu\|_V\rightarrow 0$ as $n\uparrow\infty$ fails to hold
too. In light of Theorem \ref{thm1}, we have failure of
differential entropy convergence due to failure of the
corresponding convergence for the Kullback-Liebler discriminant,
due in turn and in light of Remark \ref{Pinsker} to the respective
failure of convergence in variation.
\end{remark}

Though weak convergence does not guarantee convergence of
differential entropy, the stronger convergence in variation does
it indeed under an appropriate boundedness condition on the
densities involved. The result is the following.

\begin{theorem}\label{thmve}
Let $\{\mu_n\}_{n=1}^{\infty}\subseteq\D_+(\X)$ and
$\mu\in\D_+(\X)$ be such that $\log[\frac{d\mu}{dx}]\in
L^{\infty}(dx)$ and
$\{\log[\frac{d\mu_n}{dx}]\}_{n=1}^{\infty}\subseteq
L^{\infty}(dx)$. Assume that
\begin{equation}\label{mf}
    M\doteq\sup_{n\in\{1,2,\ldots\}}\left\|\log\left[\frac{d\mu_n}{dx}\right]\right\|_{L^{\infty}(dx)}<\infty.
\end{equation}
Then, $\{\mu_n\}_{n=1}^{\infty}\subseteq\h(\X)\cap\h(\X||\mu)$,
$\mu\in\h(\X)$ and, if $\|\mu_n-\mu\|_V\rightarrow 0$ as
$n\uparrow\infty$, we have both
\begin{equation*}
    \d[\mu_n||\mu]\rightarrow 0 \text{ and
    }\H[\mu_{n}]\rightarrow\H[\mu],
\end{equation*}
as $n\uparrow\infty$ as well.
\end{theorem}

\begin{proof}
First, since both
\begin{equation*}
    \log\left[\frac{d\mu}{dx}\right]\in L^{\infty}(dx)
\end{equation*}
and
\begin{equation*}
    \left\{\log\left[\frac{d\mu_n}{dx}\right]\right\}_{n=1}^{\infty}\subseteq
    L^{\infty}(dx),
\end{equation*}
we have $\mu\in\h(\X)$ and
$\{\mu_n\}_{n=1}^{\infty}\subseteq\h(\X)$. Therefore, we may write
\begin{align}\label{cve}
\notag    |\H[\mu_n]-\H[\mu]|&=\left|\int_{\X}\log\left[\frac{d\mu_n}{dx}\right]d\mu_n-\int_{\X}\log\left[\frac{d\mu}{dx}\right]d\mu\right|\\
\notag    &=\left|\int_{\X}\left(\frac{d\mu_n}{dx}\log\left[\frac{d\mu_n}{dx}\right]-\frac{d\mu}{dx}\log\left[\frac{d\mu}{dx}\right]\right)dx\right|\\
    &\leq\int_{\X}\left|\frac{d\mu_n}{dx}\log\left[\frac{d\mu_n}{dx}\right]-\frac{d\mu}{dx}\log\left[\frac{d\mu}{dx}\right]\right|dx.
\end{align}
Now, for each $n\in\{1,2,\ldots\}$ we have
\begin{align}\label{cve2}
    \left|\frac{d\mu_n}{dx}(x)\log\left[\frac{d\mu_n}{dx}(x)\right]-\frac{d\mu}{dx}(x)\log\left[\frac{d\mu}{dx}(x)\right]\right|
    \leq&\left|\log\left[\frac{d\mu_n}{dx}(x)\right]\right|\left|\frac{d\mu_n}{dx}(x)-\frac{d\mu}{dx}(x)\right|\nonumber\\
    &+\frac{d\mu}{dx}(x)\left|\log\left[\frac{d\mu_n}{d\mu}(x)\right]\right|,
\end{align}
for Lebesgue-almost every $x\in\X$. For each $n\in\{1,2,\ldots\}$
we also have, for Lebesgue-almost every $x\in\X$ as well,
\begin{equation}\label{af1}
    \left|\log\left[\frac{d\mu_n}{d\mu}(x)\right]\right|\leq\frac{\log\left[M'\right]}{M'-1}\left|\frac{d\mu_n}{d\mu}(x)-1\right|,
\end{equation}
since
\begin{align*}
\frac{d\mu_n}{d\mu}(x)&=\frac{d\mu_n}{dx}(x)\left[\frac{d\mu}{dx}(x)\right]^{-1}\\
            &\geq 2^{-M}2^{-\left\|\frac{d\mu}{dx}\right\|_{L^{\infty}(dx)}}\\
    &=2^{-\left(M+\left\|\frac{d\mu}{dx}\right\|_{L^{\infty}(dx)}\right)}\doteq M'\in(0,1)
\end{align*}
for each $n\in\{1,2,\ldots\}$ and Lebesgue-almost every $x\in\X$
too, and
\begin{equation}\nonumber
    |\log[a]|\leq\frac{\log[a_0]}{a_0-1}\left|a-1\right|
\end{equation}
for all $a\in[a_0,\infty)$, with $a_0\in(0,1)$. We also have
\begin{equation}\label{af2}
    \left|\frac{d\mu_n}{d\mu}(x)-1\right|=\left[\frac{d\mu}{dx}(x)\right]^{-1}\left|\frac{d\mu_n}{dx}(x)-\frac{d\mu}{dx}(x)\right|
\end{equation}
for each $n\in\{1,2,\ldots\}$ and Lebesgue-almost every $x\in\X$.
Hence, from equations (\ref{cve2}), (\ref{af1}) and (\ref{af2}) we
conclude
\begin{equation}
    \left|\frac{d\mu_n}{dx}(x)\log\left[\frac{d\mu_n}{dx}(x)\right]-\frac{d\mu}{dx}(x)\log\left[\frac{d\mu}{dx}(x)\right]\right|\\
    \leq
    \left[M+\frac{\log\left[M'\right]}{M'-1}\right]\left|\frac{d\mu_n}{dx}(x)-\frac{d\mu}{dx}(x)\right|,
\end{equation}
for each $n\in\{1,2,\ldots\}$ and Lebesgue-almost every $x\in\X$
as well, and therefore, from equation (\ref{cve}),
\begin{equation}\label{af3}
    |\H[\mu_n]-\H[\mu]|\leq
    \left[M+\frac{\log\left[M'\right]}{M'-1}\right]\left\|\frac{d\mu_n}{dx}-\frac{d\mu}{dx}\right\|_{L^1(dx)}.
\end{equation}
In the same way, since
\begin{equation}\nonumber
    \int_{\X}\left|\log\left[\frac{d\mu_n}{d\mu}\right]\right|d\mu_n
    \leq M+\left\|\log\left[\frac{d\mu}{dx}\right]\right\|_{L^{\infty}(dx)}<\infty,
\end{equation}
and therefore $\{\mu_n\}_{n=1}^{\infty}\subseteq\h(\X||\mu)$, from
equations (\ref{af1}) and (\ref{af2}) it is easy to see that
\begin{equation}\label{af4}
    \d[\mu_n||\mu]
    \leq \frac{\log\left[M'\right]}{M'\left(M'-1\right)}
    \left\|\frac{d\mu_n}{dx}-\frac{d\mu}{dx}\right\|_{L^1(dx)}.
\end{equation}
The last part of the theorem then follows from equations
(\ref{af3}) and (\ref{af4}) since, if $\|\mu_n-\mu\|_V\rightarrow
0$ as $n\uparrow\infty$, then
\begin{equation}\nonumber
    \left\|\frac{d\mu_n}{dx}-\frac{d\mu}{dx}\right\|_{L^1(dx)}\left(=\|\mu_n-\mu\|_V\right)\rightarrow0,
\end{equation}
as $n\uparrow\infty$ as well.
\end{proof}

\begin{remark}
The reader can verify that the arguments leading to the proof of
Theorem \ref{thmve} require for the supports of the densities
$\frac{d\mu}{dx}$ and $\frac{d\mu_n}{dx}$, $n\in\{1,2,\ldots\}$,
when regarded as densities in $\R^k$, to at most pairwise differ
by a Lebesgue-null set. The set $\X$ in the statement of the
theorem can then be taken as the intersection of all the afore
mentioned supports. Indeed, for such a $\mu\in\D(\R^k)$ and
$\{\mu_n\}_{n=1}^{\infty}\subseteq\D(\R^k)$ we have, with
$\mu_0\doteq\mu$, $K_0\doteq\supp(\frac{d\mu_0}{dx})$,
$K_n\doteq\supp(\frac{d\mu_n}{dx})$ for each $n\in\{1,2,\ldots\}$
(all the supports being taken w.r.t. $\R^k$) and
\begin{equation}\nonumber
    \X\doteq\bigcap_{n=0}^{\infty}K_n,
\end{equation}
that for each $m\in\{0,1,2,\ldots\}$
\begin{equation}\nonumber
    K_m\setminus\X=\bigcup_{n=0}^{\infty}\left(K_m\setminus
    K_n\right),
\end{equation}
a Lebesgue-null set, and therefore, since
$\{\mu_n\}_{n=0}^{\infty}\subseteq\D(\R^k)$, that each element in
the sequence $\{\mu_n\}_{n=0}^{\infty}$ is concentrated on $\X$.
Moreover, $\{\mu_n\}_{n=0}^{\infty}\subseteq\D_+(\X)$.
\end{remark}

We have the following corollary to Theorems \ref{thm1} and
\ref{thmve}.

\begin{corollary}\label{cuc}
Let $\{\mu_n\}_{n=1}^{\infty}\subseteq\D_+(\X)$ and $\mu\in\D(\X)$
be such that $\frac{d\mu}{dx}(x)>0$, for each $x\in\X$,
$\log[\frac{d\mu}{dx}]\in\mathcal{C}_b(\X)$ and
$\{\log[\frac{d\mu_n}{dx}]\}_{n=1}^{\infty}\subseteq
L^{\infty}(dx)$. Assume that
\begin{equation}\nonumber
    \sup_{n\in\{1,2,\ldots\}}\left\|\log\left[\frac{d\mu_n}{dx}\right]\right\|_{L^{\infty}(dx)}<\infty.
\end{equation}
Then, $\{\mu_n\}_{n=1}^{\infty}\subseteq\h(\X)\cap\h(\X||\mu)$,
$\mu\in\h(\X)$ and the following assertions are equivalent.
\begin{enumerate}
    \item $\mu_n\Rightarrow\mu$ and $\H[\mu_n]\rightarrow\H[\mu]$ as
    $n\uparrow\infty$.
    \item $\d[\mu_n||\mu]\rightarrow 0$ as $n\uparrow\infty$.
    \item$\|\mu_n-\mu\|_V\rightarrow 0$ as $n\uparrow\infty$.
\end{enumerate}
\end{corollary}

\begin{proof}
The result follows from Theorems \ref{thm1} and \ref{thmve}, in
view of Remark \ref{Pinsker} and Lemma \ref{VimpliesW}.
\end{proof}

\section{Pointwise Convergence and Differential Entropy Convergence}
\label{general}

In this section we provide a general result for convergence of
Shannon Differential Entropy, and Kullback-Liebler discriminant as
well, under an appropriate pointwise convergence condition. We
take into account both compactly and uncomplactly supported
densities. As mentioned in Section \ref{intro}, the proof is based
on exploiting absolute continuity properties of measures, in
conjunction with a suitable boundedness condition and the
dominated convergence theorem. The result is the following.

\begin{theorem}\label{thm2}
Let $\mu\in\h(\X)$ and
$\{\mu_{n}\}_{n=1}^{\infty}\subseteq\D(\X||\mu)$ be such that
$\frac{d\mu_n}{d\mu}(x)\rightarrow 1$ as $n\uparrow\infty$, for
$\mu$-almost every $x\in\X$, and
$\{\frac{d\mu_n}{d\mu}\}_{n=1}^{\infty}\subseteq
L^{\infty}(d\mu)$. Assume that
\begin{equation}\label{pur}
    M\doteq\sup_{n\in\{1,2,\ldots\}}\left\|\frac{d\mu_n}{d\mu}\right\|_{L^{\infty}(d\mu)}<\infty.
\end{equation}
Then, $\{\mu_n\}_{n=1}^{\infty}\subseteq\h(\X)\cap\h(\X||\mu)$ and
we have both
\begin{equation*}
\d[\mu_n||\mu]\rightarrow 0 \text{ and }\H[\mu_{n}]\rightarrow\H[\mu]
\end{equation*}
as $n\uparrow\infty$.
\end{theorem}

\begin{proof}
First, for each $n\in\{1,2,\ldots\}$ we have
\begin{align*}
    \int_{\X}\left|\log\left[\frac{d\mu}{dx}\right]\right|d\mu_n&=
    \int_{\X}\left|\log\left[\frac{d\mu}{dx}\right]\right|\frac{d\mu_n}{d\mu}d\mu\\
    &\leq M\int_{\X}\left|\log\left[\frac{d\mu}{dx}\right]\right|d\mu \\
     &<\infty
\end{align*}
($\mu\in\h(\X)$). Condition (\ref{pur}) in the statement of the
theorem also implies that
$\{\frac{d\mu_n}{d\mu}\log[\frac{d\mu_n}{d\mu}]\}_{n=1}^{\infty}\subseteq
L^{\infty}(d\mu)$ with
\begin{equation}\label{ua}
    M'\doteq\sup_{n\in\{1,2,\ldots\}}\left\|\frac{d\mu_n}{d\mu}\log\left[\frac{d\mu_n}{d\mu}\right]\right\|_{L^{\infty}(d\mu)}<\infty,
\end{equation}
and therefore,
$\{\mu_n\}_{n=1}^{\infty}\subseteq\h(\X||\mu)$. Indeed, for each
$n\in\{1,2,\ldots\}$,
\begin{equation}\label{fin}
   \int_{\X}\left|\log\left[\frac{d\mu_n}{d\mu}\right]\right|d\mu_{n}
    =\int_{\X}\left|\log\left[\frac{d\mu_n}{d\mu}\right]\right|\frac{d\mu_{n}}{d\mu}d\mu
    \leq\int_{\X} M'd\mu,
\end{equation}
and $\int_{\X}M'd\mu=M'\mu(\X)=M'<\infty$. Hence, for each
$n\in\{1,2,\ldots\}$ we also have
\begin{align*}
\int_{\X}\left|\log\left[\frac{d\mu_n}{dx}\right]\right|d\mu_n&\leq\int_{\X}\left|\log\left[\frac{d\mu_n}{d\mu}\right]\right|d\mu_n
    +\int_{\X}\left|\log\left[\frac{d\mu}{dx}\right]\right|d\mu_n \\
    &<\infty,
\end{align*}
thus $\{\mu_n\}_{n=1}^{\infty}\subseteq\h(\X)$, and we may write
\begin{align*}
|\H[\mu_{n}]-\H[\mu]|=&\left|\int_{\X}\log\left[\frac{d\mu_n}{dx}\right]d\mu_{n}-\int_{\X}\log\left[\frac{d\mu}{dx}\right]d\mu\right|\\  \leq&\left|\int_{\X}\log\left[\frac{d\mu_n}{dx}\right]d\mu_n-\int_{\X}\log\left[\frac{d\mu}{dx}\right]d\mu_{n}\right|\\
    &+\left|\int_{\X}\log\left[\frac{d\mu}{dx}\right]d\mu_{n}
    -\int_{\X}\log\left[\frac{d\mu}{dx}\right]d\mu\right|,
\end{align*}
i.e.,
\begin{equation}\label{maineq}
    |\H[\mu_{n}]-\H[\mu]|
    \leq\ \ \d[\mu_n||\mu]\\
    +\left|\int_{\X}\log\left[\frac{d\mu}{dx}\right]d\mu_{n}
    -\int_{\X}\log\left[\frac{d\mu}{dx}\right]d\mu\right|,
\end{equation}
for each $n\in\{1,2,\ldots\}$ as well. But,
\begin{equation}\label{KL}
    \d[\mu_n||\mu]=\int_{\X}\log\left[\frac{d\mu_n}{d\mu}\right]d\mu_n
    =\int_{\X}\log\left[\frac{d\mu_n}{d\mu}\right]\frac{d\mu_{n}}{d\mu}d\mu
\end{equation}
for each $n\in\{1,2,\ldots\}$ and, as already used in equation
(\ref{fin}), from (\ref{ua}) it follows that, for each
$n\in\{1,2,\ldots\}$ too,
\begin{equation}\nonumber
    \frac{d\mu_{n}}{d\mu}(x)\left|\log\left[\frac{d\mu_n}{d\mu}(x)\right]\right|\leq M'
\end{equation}
for $\mu$-almost every $x\in\X$. Since also
$\{\frac{d\mu_n}{d\mu}\log[\frac{d\mu_n}{d\mu}]\}_{n=1}^{\infty}$
converges pointwise $\mu$-almost everywhere to $\mathbf{0}$ on
$\X$ as $n\uparrow\infty$, where $\mathbf{0}(x)\doteq 0$,
$x\in\X$, by Lebesgue's Dominated Convergence Theorem (see for
example \cite{R1988}) we conclude
\begin{equation}\label{tf2}
    \int_{\X}\log\left[\frac{d\mu_n}{d\mu}\right]\frac{d\mu_{n}}{d\mu}d\mu\rightarrow
    0,
\end{equation}
as $n\uparrow\infty$ as well. The claimed convergence
$\d[\mu_n||\mu]\rightarrow 0$ as $n\uparrow\infty$ then follows
from equations (\ref{KL}) and (\ref{tf2}). Now, to establish the
remaining claimed convergence $\H[\mu_n]\rightarrow\H[\mu]$ as
$n\uparrow\infty$, we note that for each $n\in\{1,2,\ldots\}$ we
also have
\begin{align}\label{cuoc4}
\notag  \left|\int_{\X}\log\left[\frac{d\mu}{dx}\right]d\mu_{n}
\right. &-\left.\int_{\X}\log\left[\frac{d\mu}{dx}\right]d\mu\right|\\
\notag    &=\left|\int_{\X}\log\left[\frac{d\mu}{dx}\right]\frac{d\mu_{n}}{d\mu}d\mu-\int_{\X}\log\left[\frac{d\mu}{dx}\right]d\mu\right|\\
&\leq\int_{\X}\left|\log\left[\frac{d\mu}{dx}\right]\left(\frac{d\mu_n}{d\mu}-\mathbf{1}\right)\right|d\mu
\end{align}
(recall $\mathbf{1}(x)\doteq 1$, $x\in\X$). But, since
$\{\frac{d\mu_n}{d\mu}\}_{n=1}^{\infty}$ converges pointwise
$\mu$-almost everywhere to $\mathbf{1}$ on $\X$ as
$n\uparrow\infty$, we conclude
\begin{equation}\nonumber
    \log\left[\frac{d\mu}{dx}\right]\left(\frac{d\mu_n}{d\mu}-\mathbf{1}\right)\rightarrow\mathbf{0},
\end{equation}
$\mu$-almost everywhere on $\X$ and as $n\uparrow\infty$ as well.
In addition, since we obviously also have
$\{\frac{d\mu_{n}}{d\mu}-\mathbf{1}\}_{n=1}^{\infty}\subseteq
L^{\infty}(d\mu)$ and
\begin{align*}
    M''&\doteq\sup_{n\in\{1,2,\ldots\}}\left\|\frac{d\mu_{n}}{d\mu}-\mathbf{1}\right\|_{L^{\infty}(d\mu)}\\
&\leq\sup_{n\in\{1,2,\ldots\}}\left\|\frac{d\mu_{n}}{d\mu}\right\|_{L^{\infty}(d\mu)}+1 \\
&=M+1 \\
&<\infty,
\end{align*}
we conclude that, for each $n\in\{1,2,\ldots\}$,
\begin{equation}\nonumber
    \left|\log\left[\frac{d\mu}{dx}(x)\right]\left(\frac{d\mu_n}{d\mu}(x)-1\right)\right|\leq
    M''\left|\log\left[\frac{d\mu}{dx}(x)\right]\right|
\end{equation}
for $\mu$-almost every $x\in\X$. But,
\begin{equation}\nonumber
    \int_{\X}M''\left|\log\left[\frac{d\mu}{dx}\right]\right|d\mu=M''\int_{\X}\left|\log\left[\frac{d\mu}{dx}\right]\right|d\mu<\infty
\end{equation}
($\mu\in\h(\X)$). Hence, once again by Lebesgue's Dominated
Convergence Theorem we conclude
\begin{equation}\nonumber
    \int_{\X}\left|\log\left[\frac{d\mu}{dx}\right]\left(\frac{d\mu_n}{d\mu}-\mathbf{1}\right)\right|d\mu\rightarrow 0
\end{equation}
as $n\uparrow\infty$, and therefore from equation (\ref{cuoc4}) we
also have
\begin{equation}\label{tf3}
    \int_{\X}\log\left[\frac{d\mu}{dx}\right]d\mu_{n}\rightarrow\int_{\X}\log\left[\frac{d\mu}{dx}\right]d\mu,
\end{equation}
as $n\uparrow\infty$ as well. The claimed convergence
$\H[\mu_{n}]\rightarrow\H[\mu]$ as $n\uparrow\infty$ now follows
from equations (\ref{maineq}), (\ref{KL}), (\ref{tf2}) and
(\ref{tf3}), proving the theorem.
\end{proof}

\begin{remark}\label{re}
If $\{\mu_n\}_{n=1}^{\infty}\subseteq\D(\X)$ and
$\mu\in\D_+(\X)$, then $\mu$-almost everywhere pointwise
convergence $\frac{d\mu_n}{d\mu}\rightarrow \mathbf{1}$ as
$n\uparrow\infty$ is equivalent to Lebesgue-almost everywhere
pointwise convergence
$\frac{d\mu_n}{dx}\rightarrow\frac{d\mu}{dx}$, as
$n\uparrow\infty$ as well (both on $\X$, of course).
\end{remark}

\begin{remark}\label{PCCV}
If $\{\mu_{n}\}_{n=1}^{\infty}\subseteq\D(\X)$ and
$\mu\in\D(\X)$ with $\{\frac{d\mu_n}{dx}\}_{n=1}^{\infty}$
converging pointwise Lebesgue-almost everywhere to
$\frac{d\mu}{dx}$ on $\X$ as $n\uparrow\infty$, then
$\|\mu_n-\mu\|_V\rightarrow 0$, as $n\uparrow\infty$ as well.
Indeed,
\begin{equation}\nonumber
    \|\mu_n-\mu\|_V=
    \left\|\frac{d\mu_n}{dx}-\frac{d\mu}{dx}\right\|_{L^1(dx)}
    =\int_{\X}\left|\frac{d\mu_n}{dx}-\frac{d\mu}{dx}\right|dx\rightarrow 0
\end{equation}
as $n\uparrow\infty$, the convergence following from Scheff\'{e}'s
Lemma, \cite[Lemma 5.10, p.55]{DW1991}. Therefore, when
$\{\mu_n\}_{n=1}^{\infty}\subseteq\D_+(\X)$ and $\mu\in\D_+(\X)$,
by going from convergence in variation in Theorem \ref{thmve}, to
pointwise convergence of the corresponding densities in Theorem
\ref{thm2} (see Remark \ref{re} above), we are able to relax the
corresponding boundedness condition from (\ref{mf}) to
(\ref{pur}). Indeed, for
$\{\mu_n\}_{n=1}^{\infty}\subseteq\D_+(\X)$ and $\mu\in \D_+(\X)$
satisfying $\log[\frac{d\mu}{dx}]\in L^{\infty}(dx)$ and
$\{\log[\frac{d\mu_n}{dx}]\}_{n=1}^{\infty}\subseteq
L^{\infty}(dx)$ with
$\sup_{n\in\{1,2,\ldots\}}\|\log[\frac{d\mu_n}{dx}]\|_{L^{\infty}(dx)}<\infty$,
we have $\{\frac{d\mu_n}{d\mu}\}_{n=1}^{\infty}\subseteq
L^{\infty}(d\mu)$ and
\begin{equation}\nonumber
    \sup_{n\in\{1,2,\ldots\}}\left\|\frac{d\mu_n}{d\mu}\right\|_{L^{\infty}(d\mu)}\leq
    \frac{2^{\sup_{n\in\{1,2,\ldots\}}
    \left\|\log\left[\frac{d\mu_n}{dx}\right]\right\|_{L^{\infty}(dx)}}}{2^{-\left\|\frac{d\mu}{dx}\right\|_{L^{\infty}(dx)}}},
\end{equation}
condition (\ref{mf}) implying then (\ref{pur}).
\end{remark}

\begin{remark}
For any $\{\mu_{n}\}_{n=1}^{\infty}\subseteq\D(\X)$ and
$\mu\in\D_+(\X)$, condition (\ref{pur}) in Theorem \ref{thm2}
reads as
\begin{equation}\label{leb}
    \frac{d\mu_n}{dx}(x)\left[\frac{d\mu}{dx}(x)\right]^{-1}\leq
    M<\infty
\end{equation}
for each $n\in\{1,2,\ldots\}$ and Lebesgue-almost every $x\in\X$,
and therefore, as the reader can easily verify (note $M\geq 1$
necessarily), we have
\begin{equation}\nonumber
    \frac{d\mu_n}{dx}(x)\log\left[\frac{d\mu_n}{dx}(x)\right]
    \leq
    M\max\left\{\psi_1(x),\psi_2(x)\right\},
\end{equation}
for each $n\in\{1,2,\ldots\}$ and Lebesgue-almost every $x\in\X$
as well, where
\begin{equation}\nonumber
    \psi_1(x)\doteq\frac{d\mu}{dx}(x)\log\left[M\right]
\end{equation}
and
\begin{equation}\nonumber
    \psi_2(x)\doteq\frac{d\mu}{dx}(x)\log\left[M\right]+\frac{d\mu}{dx}(x)\log\left[\frac{d\mu}{dx}(x)\right].
\end{equation}
Thus, since also $y\log[y]\geq(-e\ln[2])^{-1}$ for all $y\in\R_+$
(recall $0\log[0]=0[-\infty]=0$ by convention), condition
(\ref{leb}) then implies the existence of $C_0, C_1, C_2\in\R_+$,
with $C_0>0$ necessarily if $\X$ has infinite Lebesgue measure
(easy to check), such that for each $n\in\{1,2,\ldots\}$
\begin{equation}\label{leb2}
    \left|\frac{d\mu_n}{dx}(x)\log\left[\frac{d\mu_n}{dx}(x)\right]\right|
    \leq f(x),
\end{equation}
for Lebesgue-almost every $x\in\X$, where
\begin{equation}\nonumber
    f(x)\doteq C_0+C_1\frac{d\mu}{dx}(x)+C_2\frac{d\mu}{dx}(x)\left|\log\left[\frac{d\mu}{dx}(x)\right]\right|.
\end{equation}
However, even with $\{\mu_n\}_{n=1}^{\infty}\subseteq\h(\X)$,
$\mu\in\h(\X)\cap\D_+(\X)$ and
$\{\frac{d\mu_n}{dx}\}_{n=1}^{\infty}$ converging pointwise
Lebesgue-almost everywhere to $\frac{d\mu}{dx}$ on $\X$ as
$n\uparrow\infty$ (see Remark \ref{re}), condition (\ref{leb2})
cannot be used in the dominated convergence theorem to conclude
the convergence
\begin{align*}
    \H\left[\mu_{n}\right]&=-\int_{\X}\log\left[\frac{d\mu_n}{dx}\right]d\mu_n\\
    &=-\int_{\X}\log\left[\frac{d\mu_n}{dx}\right]\frac{d\mu_n}{dx}dx\\
    &\rightarrow
    -\int_{\X}\log\left[\frac{d\mu}{dx}\right]\frac{d\mu}{dx}dx\\
    &=-\int_{\X}\log\left[\frac{d\mu}{dx}\right]d\mu\\
    &=\H\left[\mu\right],
\end{align*}
as $n\uparrow\infty$, if $\X$ has infinite Lebesgue measure.
Indeed,
\begin{equation}\nonumber
    \int_{\X}fdx\geq\int_{\X}C_0dx=C_0\int_{\X}dx=\infty
\end{equation}
in that case. Therefore the advantage of considering integrals
w.r.t. $d\mu$ (instead of $dx$) in the arguments leading to the
proof of Theorem \ref{thm2}.
\end{remark}

\section{Discrete Alphabet Sources}
\label{discrete}

In this section we consider discrete alphabet sources. We show how
all convergence results become straightforward for finitely
supported probability measures, and we also provide results for
the infinitely supported case, by exploiting the equivalence
between weak convergence and convergence in variation in this
setting.

Though most of the definitions in the previous sections include
the discrete case as a particular case when no reference to
$\D(\X)$ is made, by considering then discretely supported
probability measures, for sake of preciseness we briefly go
through all the relevant concepts before stating the results.

Throughout this section we consider, specifically,
\begin{equation*}
\X\doteq\{x_i\}_{i\in\I}\subseteq\R^k \text{ with }
\I\subseteq\{1,2,\ldots\}
\end{equation*}
and, as before, $\R^k$ the $k$-dimensional Euclidian space. (Note
that $\I$ is allowed to be the whole of $\{1,2,\ldots\}$.)
Accordingly, $\mathcal{S}(\X)$ denotes the collection of all
subsets of $\X$ and $\P(\X)$ the collection of all probability
measures on $(\X,\mathcal{S}(\X))$. A measure $\mu\in\P(\X)$ is
now characterized by the sequence
$\{p_i^{\mu}\}_{i\in\I}\subseteq[0,1]$, satisfying the
normalization condition
\begin{equation*}
\sum_{i\in\I}p_i^{\mu}=1,
\end{equation*}
given by $p_i^{\mu}\doteq\mu(\{x_i\})$, $i\in\I$. To any sequence
$\{a_i\}_{i\in\I}\subseteq\R$ we associate the mapping
$a:\X\rightarrow\R$ by setting $a(x_i)\doteq a_i$ for each
$i\in\I$. We shall use the same notation as in the previous
sections to denote now (recall the conventions $\log[0]=-\infty$
and $0[\pm\infty]=0$)
\begin{equation}\nonumber
   \h(\X)\doteq\left\{\mu\in\P(\X):
   \left\{\log\left[p_i^{\mu}\right]\right\}_{i\in\I}\in
   l^{1}(\mu)\right\},
\end{equation}
where
\begin{equation}\nonumber
    l^{1}(\mu)\doteq\left\{\{a_i\}_{i\in\I}\subseteq\R:\sum_{i\in\I}|a_i|p_i^{\mu}<\infty\right\}
\end{equation}
for $\mu\in\P(\X)$. Note that $\{a_i\}_{i\in\I}\in l^{1}(\mu)$ if
and only if $a\in L^{1}(d\mu)$. In fact,
\begin{equation*}
\sum_{i\in\I}a_ip_i^{\mu}=\int_{\X}ad\mu
\end{equation*}
and
\begin{equation*}
    \left\|\{a_i\}_{i\in\I}\right\|_{l^1(\mu)}\doteq\sum_{i\in\I}\left|a_i\right|p_i^{\mu}=\int_{\X}\left|a\right|d\mu=\left\|a\right\|_{L^{1}(d\mu)}
\end{equation*}
for $\{a_i\}_{i\in\I}\in l^{1}(\mu)$ or, equivalently, for $a\in
L^1(d\mu)$.

\begin{remark}
For any given $\mu\in\P(\X)$, Banach spaces
($l^p(\mu),\|\cdot\|_{l^p(\mu)})$ can be considered for each
$p\in[1,\infty]$, similarly than in Section \ref{preliminary},
provided sequences coinciding at each $i\in\I$ for which
$p^{\mu}_i>0$ (i.e., $\mu$-almost everywhere) are treated as
equivalent.
\end{remark}

Consider the measure $\mu\in\h(\X)$. Then,
\begin{equation}\nonumber
   \H[\mu]\doteq-\sum_{i\in\I}p_i^{\mu}\log\left[p_i^{\mu}\right]\in\R_+
\end{equation}
is the Shannon Entropy of $\mu$. Also, given $\mu\in\P(\X)$,
$\D(\X||\mu)$ denotes the set of all probability measures
$\sigma\in\P(\X)$ that are absolutely continuous w.r.t. $\mu$,
i.e., satisfying the condition $p_i^{\sigma}=0$ whenever
$p_i^{\mu}=0$. In the same way,
\begin{equation}\nonumber
    \h(\X||\mu)\doteq\left\{\sigma\in\D(\X||\mu):
    \left\{\log\left[\frac{p_i^{\sigma}}{p_i^{\mu}}\right]\right\}_{i\in\I}\in l^{1}(\sigma)\right\},
\end{equation}
with the standard convention $0\log[\frac{0}{0}]=0$ (motivated by
continuity). Finally, if $\sigma\in\h(\X||\mu)$, then
\begin{equation}\nonumber
    \d[\sigma||\mu]\doteq\sum_{i\in\I}p_i^{\sigma}\log\left[\frac{p_i^{\sigma}}{p_i^{\mu}}\right]\in\R_+
\end{equation}
is the Shannon Relative Entropy between $\sigma$ and $\mu$, or
equivalently the Kullback-Liebler Discriminant between $\sigma$
and $\mu$ too. It is worthy to emphasize that
\begin{equation*}
\sum_{i\in\I}p_i^{\sigma}\log\left[\frac{p_i^{\sigma}}{p_i^{\mu}}\right]\geq
0
\end{equation*}
for any $\mu\in\P(\X)$ and $\sigma\in\D(\X||\mu)$, with equality
if and only if $p^{\mu}=p^{\sigma}$ $\sigma$-almost everywhere,
i.e., if and only if $p_i^{\mu}=p_i^{\sigma}$ for each $i\in\I$
such that $p_i^{\sigma}>0$.

Weak convergence of $\{\mu_n\}_{n=1}^{\infty}\subseteq\P(\X)$ to
$\mu\in\P(\X)$ is now characterized as follows. We have
$\mu_n\Rightarrow\mu$ as $n\uparrow\infty$ if and only if
\begin{equation}\label{WCD}
    \sum_{i\in\I}f(x_i)p_i^{\mu_n}\rightarrow\sum_{i\in\I}f(x_i)p_i^{\mu},
\end{equation}
as $n\uparrow\infty$ as well, for each bounded, real-valued
function $f$ on $\X$.

Distance in variation between $\sigma_1\in\P(\X)$ and
$\sigma_2\in\P(\X)$ is
\begin{equation}\nonumber
    \|\sigma_1-\sigma_2\|_V=\left\|\left\{p_i^{\sigma_1}-p_i^{\sigma_2}\right\}_{i\in\I}\right\|_{l^{1}(\delta)},
\end{equation}
where $\delta$ denotes the counting measure on
$(\X,\mathcal{S}(\X))$, i.e., $\delta(\{x_i\})\doteq 1$ for each
$i\in\I$ and $\delta(A)\doteq\sum_{x_i\in A}\delta(\{x_i\})$ for
each $A\in\mathcal{S}(\X)$. (As before, note
$\{p_i^{\sigma_1}-p_i^{\sigma_2}\}_{i\in\I}\in l^1(\delta)$ for
all $\sigma_1,\sigma_2\in\P(\X)$.) The corresponding convergence
in variation of $\{\mu_n\}_{n=1}^{\infty}\subseteq\P(\X)$ to
$\mu\in\P(\X)$, $\|\mu_n-\mu\|_V\rightarrow 0$ as
$n\uparrow\infty$, takes place if and only if
\begin{equation}\nonumber
    \sum_{i\in\I}\left|p_i^{\mu_n}-p_i^{\mu}\right|\rightarrow 0,
\end{equation}
as $n\uparrow\infty$ as well, since
\begin{equation}\nonumber
    \left\|\{p_i^{\mu_n}-p_i^{\mu}\}_{i\in\I}\right\|_{l^1(\delta)}=\sum_{i\in\I}\left|p_i^{\mu_n}-p_i^{\mu}\right|.
\end{equation}

In the discrete setting, the relationship between weak convergence
and convergence in variation in Lemma \ref{VimpliesW} can be
strengthened, as stated in the following result.

\begin{lemma}\label{ViifW}
Let $\{\mu_n\}_{n=1}^{\infty}\subseteq\P(\X)$ and $\mu\in\P(\X)$.
Then, we have $\mu_n\Rightarrow\mu$ as $n\uparrow\infty$ if and
only if $\|\mu_n-\mu\|_V\rightarrow 0$, as $n\uparrow\infty$ as
well. Moreover, the previous ways of convergence take place if and
only if $p_i^{\mu_n}\rightarrow p_i^{\mu}$ as $n\uparrow\infty$
for each $i\in\I$, i.e., both the topology of weak convergence and
convergence in variation are equivalent to the topology of
coordinatewise convergence of the sequence of vectors
$\{(p_i^{\mu_n})_{i\in\I}\}_{n=1}^{\infty}$ to the vector
$(p_i^{\mu})_{i\in\I}$ as $n\uparrow\infty$ (equivalently, to the
topology of pointwise convergence of
$\{p^{\mu_n}\}_{n=1}^{\infty}$ to $p^{\mu}$ on $\X$ as
$n\uparrow\infty$).
\end{lemma}

\begin{proof}
We obviously have that $\mu_n\Rightarrow\mu$ as $n\uparrow\infty$
implies $p_i^{\mu_n}\rightarrow p_i^{\mu}$, as $n\uparrow\infty$
as well, for each $i\in\I$. Indeed, we just need to consider
equation (\ref{WCD}) with $f_i:\X\rightarrow\{0,1\}$ defined, for
each $i\in\I$, by letting $f_i(x)\doteq 1$ if $x=x_i$ and
$f_i(x)\doteq 0$ if $x\in\X\setminus\{x_i\}$. Now, since
\begin{align*}
\|\mu_n-\mu\|_V&=\left\|\left\{p_i^{\mu_n}-p_i^{\mu}\right\}_{i\in\I}\right\|_{l^{1}(\delta)}\\
&=\left\|p^{\mu_n}-p^{\mu}\right\|_{L^1(d\delta)}\\
&=\int_{\X}|p^{\mu_n}-p^{\mu}|d\delta,
\end{align*}
if the sequence $\{p^{\mu_n}\}_{n=1}^{\infty}$ converges pointwise
to $p^{\mu}$ on $\X$ as $n\uparrow\infty$, then Scheff\'{e}'s
Lemma gives us the convergence $\|\mu_n-\mu\|_V\rightarrow 0$, as
$n\uparrow\infty$ too, the same as in the differential case (see
Remark \ref{PCCV}). The lemma then follows from Lemma
\ref{VimpliesW}.
\end{proof}

\begin{remark}
In the differential setting and from Remark \ref{PCCV} and Lemma
\ref{VimpliesW}, we have the chain of implications: pointwise
convergence of densities (Lebesgue-almost everywhere pointwise
convergence in fact) $\Rightarrow$ convergence in variation
$\Rightarrow$ weak convergence. As Lemma \ref{ViifW} shows, the
corresponding three ways of convergence in the discrete setting
are indeed equivalent.
\end{remark}

In view of Lemma \ref{ViifW}, it is a straightforward exercise to
check that in the case when the set $\X$ (equivalently the index
set $\I$) can be taken to be finite (i.e., when the supports of
all probability measures involved are contained in a finite set),
the convergence $\mu_n\Rightarrow\mu$ as $n\uparrow\infty$ implies
both
\begin{equation*}
\H[\mu_n]\rightarrow\H[\mu]\text{ and } \d[\mu_n||\mu]\rightarrow 0,
\end{equation*}
as $n\uparrow\infty$ as well, being in fact, from Remark
\ref{Pinsker}, $\mu_n\Rightarrow\mu$ as $n\uparrow\infty$ and
$\d[\mu_n||\mu]\rightarrow 0$ as $n\uparrow\infty$ equivalent (of
course with $\{\mu_n\}_{n=1}^{\infty}\subseteq\D(\X||\mu)$ for the
convergence $\d[\mu_n||\mu]\rightarrow 0$ as $n\uparrow\infty$).

Given $\mu\in\P(\X)$ and
$\{\mu_n\}_{n=1}^{\infty}\subseteq\P(\X)$, the set $\X$ can be
made into a finite set whenever $\mu$ is finitely supported and
$\{\mu_n\}_{n=1}^{\infty}\subseteq\D(\X||\mu)$, by just redefining
it as $\X_{\mu}$ with
\begin{equation*}
    \X_{\mu}\doteq\supp\left(p^{\mu}\right)=\left\{x\in\X:p^{\mu}(x)>0\right\}=\{x_i\}_{i\in\I_{\mu}}
\end{equation*}
and $\I_{\mu}\doteq\{i\in\I:p_i^{\mu}>0\}$. Note that if
$\mu\in\P(\X)$ is finitely supported and
$\{\mu_n\}_{n=1}^{\infty}\subseteq\D(\X||\mu)$, then
$\mu\in\h(\X)$ and $\{\mu_n\}_{n=1}^{\infty}\subseteq\h(\X||\mu)$.
The discrete setting versions of Theorems \ref{thm1} and
\ref{thmve} and Corollary \ref{cuc} are trivial in that case. They
cannot be stated for $\mu\in\P(\X)$ being infinitely supported
however, as clear from the following remark.

\begin{remark}\label{mbf}
Unlike in the differential setting (see Remark \ref{nnfs}), in the
discrete setting we have for $\mu\in\P(\X)$ that
$p_i^{\mu}\rightarrow 0$ as $i\uparrow\infty$, $i\in\I_{\mu}$,
whenever $\I_{\mu}$ (equivalently $\X_{\mu}$) is infinite
($\sum_{i\in\I_{\mu}}p_i^{\mu}=1<\infty$), and therefore the
subsequence $\{\log[p_i^{\mu}]\}_{i\in\I_{\mu}}$ cannot be bounded
in that case (even when $\X_{\mu}$ is a bounded subset of $\R^k$).
\end{remark}

We consider the general case, covering infinitely supported
probability measures, in the following theorem (which corresponds
to the discrete setting version of Theorem \ref{thm2}) and two
corresponding corollaries. Though the proof of the theorem follows
by similar corresponding arguments as those in the proof of
Theorem \ref{thm2}, we include here the main steps in order to
make clear the connection between both settings.

\begin{theorem}\label{thmde}
Let $\mu\in\h(\X)$ and
$\{\mu_{n}\}_{n=1}^{\infty}\subseteq\D(\X||\mu)$ be such that
$p^{\mu_n}_i\rightarrow p_i^{\mu}$ as $n\uparrow\infty$, for each
$i\in\I_{\mu>0}$, and
\begin{equation}\nonumber
    M\doteq\sup_{\substack{n\in\{1,2,\ldots\}\\i\in\I_{\mu>0}}}\frac{p_i^{\mu_n}}{p_i^{\mu}}<\infty.
\end{equation}
Then, $\{\mu_n\}_{n=1}^{\infty}\subseteq\h(\X)\cap\h(\X||\mu)$ and
we have both
\begin{equation*}
    \d[\mu_n||\mu]\rightarrow 0 \text{ and }\H[\mu_{n}]\rightarrow\H[\mu]
\end{equation*}
as $n\uparrow\infty$.
\end{theorem}

\begin{proof}
As in the proof of Theorem \ref{thm2}, it is easy to see that
$\{\mu_n\}_{n=1}^{\infty}\subseteq\h(\X)\cap\h(\X||\mu)$, and that
\begin{equation}\label{thmd}
    \left|\H[\mu_n]-\H[\mu]\right|\leq\d[\mu_n||\mu]+\left|\sum_{i\in\I}\log\left[p_i^{\mu}\right]\left(p_i^{\mu_n}-p_i^{\mu}\right)\right|
\end{equation}
for each $n\in\{1,2,\ldots\}$. But, for each $n\in\{1,2,\ldots\}$
as well,
\begin{align*}
    \d[\mu_n||\mu]&=\sum_{i\in\I}p_i^{\mu_n}\log\left[\frac{p_i^{\mu_n}}{p_i^{\mu}}\right] \\
    &=\sum_{i\in\I_{\mu}}p_i^{\mu_n}\log\left[\frac{p_i^{\mu_n}}{p_i^{\mu}}\right]\\
    &=\sum_{i\in\I_{\mu}}\frac{p_i^{\mu_n}}{p_i^{\mu}}\log\left[\frac{p_i^{\mu_n}}{p_i^{\mu}}\right]p_i^{\mu}
\end{align*}
($\{\mu_{n}\}_{n=1}^{\infty}\subseteq\D(\X||\mu)$ and
$0\log[\frac{0}{0}]=0$ by convention), and
\begin{equation}\nonumber
    \sum_{i\in\I_{\mu}}\frac{p_i^{\mu_n}}{p_i^{\mu}}\log\left[\frac{p_i^{\mu_n}}{p_i^{\mu}}\right]p_i^{\mu}
    =\int_{\X_{\mu}}\frac{p^{\mu_n}}{p^{\mu}}\log\left[\frac{p^{\mu_n}}{p^{\mu}}\right]d\mu.
\end{equation}
Also, there exists $M'\in\R_+$ such that
\begin{equation*}
    \frac{p^{\mu_n}(x)}{p^{\mu}(x)}\left|\log\left[\frac{p^{\mu_n}(x)}{p^{\mu}(x)}\right]\right|\leq
    M',
\end{equation*}
for each $x\in\X_{\mu}$ and $n\in\{1,2,\ldots\}$, and
\begin{equation*}
    \int_{\X_{\mu}}M'd\mu=M'\sum_{i\in\I_{\mu}}p_i^{\mu}=M'\sum_{i\in\I}p_i^{\mu}=M'<\infty.
\end{equation*}
Then, since $p^{\mu_n}\rightarrow p^{\mu}$ pointwise on $\X_{\mu}$
as $n\uparrow\infty$, and therefore
\begin{equation*}
  \frac{p^{\mu_n}}{p^{\mu}}\log\left[\frac{p^{\mu_n}}{p^{\mu}}\right]\rightarrow\mathbf{0}
\end{equation*}
(recall $\mathbf{0}(x)\doteq 0$, $x\in\X$), pointwise on
$\X_{\mu}$ and as $n\uparrow\infty$ as well, by Lebesgue's
Dominated Convergence Theorem we conclude the convergence
\begin{equation*}
    \int_{\X_{\mu}}\frac{p^{\mu_n}}{p^{\mu}}\log\left[\frac{p^{\mu_n}}{p^{\mu}}\right]d\mu\rightarrow0
\end{equation*}
as $n\uparrow\infty$, and thus the claimed convergence
$\d[\mu_n||\mu]\rightarrow 0$ as $n\uparrow\infty$ too. Now, note
for each $n\in\{1,2,\ldots\}$ we also have
\begin{align*}
    \left|\sum_{i\in\I}\log\left[p_i^{\mu}\right]\left(p_i^{\mu_n}-p_i^{\mu}\right)\right|
    &=\left|\sum_{i\in\I_{\mu}}\log\left[p_i^{\mu}\right]\left(p_i^{\mu_n}-p_i^{\mu}\right)\right|\\
    &\leq\sum_{i\in\I_{\mu}}\left|\log\left[p_i^{\mu}\right]\left(\frac{p_i^{\mu_n}}{p_i^{\mu}}-1\right)\right|p_i^{\mu},
\end{align*}
and
\begin{equation}
    \sum_{i\in\I_{\mu}}\left|\log\left[p_i^{\mu}\right]\left(\frac{p_i^{\mu_n}}{p_i^{\mu}}-1\right)\right|p_i^{\mu}\\
    =\int_{\X_{\mu}}\left|\log\left[p^{\mu}\right]\left(\frac{p^{\mu_n}}{p^{\mu}}-\mathbf{1}\right)\right|d\mu
\end{equation}
(recall $\mathbf{1}(x)\doteq 1$, $x\in\X$). But,
\begin{equation*}
    \left|\log\left[p^{\mu}(x)\right]\left(\frac{p^{\mu_n}(x)}{p^{\mu}(x)}-1\right)\right|\leq
    M''\left(-\log\left[p^{\mu}(x)\right]\right)
\end{equation*}
for each $x\in\X_{\mu}$ and $n\in\{1,2,\ldots\}$, with $M''\doteq M+1\in\R_+$, and
\begin{align*}
    \int_{\X_{\mu}}M''\left(-\log\left[p^{\mu}\right]\right)d\mu&=M''\left(-\sum_{i\in\I_{\mu}}p_i^{\mu}\log\left[p_i^{\mu}\right]\right)\\
    &=M''\left(-\sum_{i\in\I}p_i^{\mu}\log\left[p_i^{\mu}\right]\right)\\
    &=M''\H[\mu]\\
    &<\infty.
\end{align*}
In addition, by the same arguments as before,
\begin{equation*}
    \log\left[p^{\mu}\right]\left(\frac{p^{\mu_n}}{p^{\mu}}-\mathbf{1}\right)\rightarrow\mathbf{0}
\end{equation*}
pointwise on $\X_{\mu}$ as $n\uparrow\infty$. Hence, once again by
Lebesgue's Dominated Convergence Theorem we conclude
\begin{equation*}
    \int_{\X_{\mu}}\left|\log\left[p^{\mu}\right]\left(\frac{p^{\mu_n}}{p^{\mu}}-\mathbf{1}\right)\right|d\mu\rightarrow0
\end{equation*}
as $n\uparrow\infty$, and therefore
\begin{equation}\label{thmd3}
    \sum_{i\in\I}\log\left[p_i^{\mu}\right]\left(p_i^{\mu_n}-p_i^{\mu}\right)\rightarrow 0
\end{equation}
as $n\uparrow\infty$ too. The remaining claimed convergence
$\H[\mu_n]\rightarrow\H[\mu]$ as $n\uparrow\infty$ then follows
from the convergence $\d[\mu_n||\mu]\rightarrow 0$ as
$n\uparrow\infty$ and equations (\ref{thmd}) and (\ref{thmd3}),
proving the theorem.
\end{proof}

We have the following two corollaries to Theorem \ref{thmde}. For
the first, let us define $[\h\cap\P_+](\X)\doteq\h(\X)\cap\P_+(\X)$ with
$\P_+(\X)$ the collection of all $\mu\in\P(\X)$ satisfying
$p_i^{\mu}>0$ for each $i\in\I$.

\begin{corollary}
Let $\mu\in[\h\cap\P_{+}](\X)$ and
$\{\mu_{n}\}_{n=1}^{\infty}\subseteq\P(\X)$ be such that
$\mu_n\Rightarrow\mu$ as $n\uparrow\infty$ and
\begin{equation}\nonumber
    \sup_{\substack{n\in\{1,2,\ldots\}\\i\in\I}}\frac{p_i^{\mu_n}}{p_i^{\mu}}<\infty.
\end{equation}
Then, $\{\mu_n\}_{n=1}^{\infty}\subseteq\h(\X)\cap\h(\X||\mu)$ and
we have both
\begin{equation*}
    \d[\mu_n||\mu]\rightarrow 0 \text{ and }\H[\mu_{n}]\rightarrow\H[\mu]
\end{equation*}
as $n\uparrow\infty$.
\end{corollary}

\begin{proof}
The result follows from Theorem \ref{thmde}, in view of Lemma
\ref{ViifW}.
\end{proof}

\begin{corollary}\label{cp}
Let $\mu\in\h(\X)$ and
$\{\mu_{n}\}_{n=1}^{\infty}\subseteq\D(\X||\mu)$ be such that
\begin{equation}\label{urs}
    \sup_{\substack{n\in\{1,2,\ldots\}\\i\in\I_{\mu>0}}}\frac{p_i^{\mu_n}}{p_i^{\mu}}<\infty.
\end{equation}
Then, $\{\mu_n\}_{n=1}^{\infty}\subseteq\h(\X)\cap\h(\X||\mu)$
and, if $\d[\mu_n||\mu]\rightarrow 0$ as $n\uparrow\infty$, we
have
\begin{equation*}
    \H[\mu_{n}]\rightarrow\H[\mu],
\end{equation*}
as $n\uparrow\infty$ as well.
\end{corollary}

\begin{proof}
The result follows from Theorem \ref{thmde}, in view of Remark
\ref{Pinsker} and Lemma \ref{ViifW}.
\end{proof}

\begin{remark}
In the context of continuity versus pure convergence properties of
Shannon entropy discussed in Section \ref{intro}, note Corollary
\ref{cp} establishes the convergence $\H[\mu_n]\rightarrow\H[\mu]$
as $n\uparrow\infty$, under the convergence
$\d[\mu_n||\mu]\rightarrow 0$ as $n\uparrow\infty$ as well, by
exploiting an underlying structure relating
$\{\mu_n\}_{n=1}^{\infty}$ to $\mu$ (condition (\ref{urs})). In
contrast, by imposing the stronger requirement on $\mu$ of being
power dominated (stronger than just $\mu\in\h(\X)$; see \cite{PH}
for the definition of a power dominated distribution), the
continuity result \cite[Theorem 21, p.16]{PH} establishes the
corresponding entropy convergence, in a discrete setting too, for
all approximating sequences converging in the above
Kullback-Liebler discriminant sense.
\end{remark}

\section{Conclusion}
\label{conclusions}

Results on convergence of Shannon entropy have been established
for both the differential and discrete settings. In the
differential case, it was shown that weak convergence of the
underlying probability measures is not enough for convergence of
the associated differential entropies. Differential entropy
convergence was then established for fairly general supported
densities in terms of the Kullback-Liebler discriminant, and it
was also shown that under an appropriate boundedness condition,
the stronger convergence in variation of the underlying
probability measures does indeed guarantee the desired
differential entropy convergence. A general result for
differential entropy convergence was also provided in terms of a
pointwise convergence condition, accounting for compactly and
uncompactly supported densities. In the discrete case, it was
shown that convergence in distribution and in variation of
probability measures become equivalent, trivially guaranteeing all
information measures convergence in the finitely supported case.
Results on entropy and Kullback-Liebler discriminant convergence
were also established in this setting for possibly infinite
supported probability measures.

We believe the results here exposed will find a wide scope of
applicability, specially in light of the great generality allowed
for the support sets of the probability measures involved.

\bibliographystyle{IEEEtran}
\bibliography{IEEEfull,mybib}

\end{document}